Comment on

"Rubric-based holistic review represents a change

from traditional graduate admissions approaches in physics "

M. B. Weissman

*Department of Physics, University of Illinois at Urbana-Champaign*

*1110 West Green Street, Urbana, IL 61801-3080*

Abstract: A recent paper evaluating a new rubric-based graduate admissions approach using generic methods tentatively suggested that its decisions differed noticeably from the previous approach in an unspecified way. Using prior knowledge that the often-stated specific goal was to open a path to increased diversity by reducing barriers to admission of applicants with low undergraduate grade point averages and GRE scores allows simple statistical tests of changes in the distributions of the metrics. The simple tests confirm with good statistical confidence that the barrier-reduction changes were achieved. Nevertheless, the paper's argument that the de-emphasized test and grades are not predictive of graduate outcomes is not supported by the prior literature. On technique, although a  method used in some of the analysis for dropping data points before running the machine algorithm is likely to bias those results, it helps to clarify why models of the rubric-based system were only weakly predictive.

Young et al. (*1*) have recently examined changes in graduate physics admissions results caused by switching to a new rubric-based system at Michigan State University. They initially outline the broad goal of their work as follows:

> Only three out of five physics students who enroll in a Ph.D. program will successfully complete their program…. Solutions must consider both the admission and retention sides ….we will focus on the former.

Thus the stated aim of this particular paper is to examine changes of graduate admissions procedures designed to increase graduation rates.

Although the connection to the initially stated goal of increasing graduation rates is not spelled out, a key motivation for the change to rubric-based admissions is clearly laid out in many of their references and prior publications, e.g. (*2-5*). That goal is to obtain domestic students more representative of the demographics of the United States. It is possible that new admissions procedures would achieve both goals. Rubrics are intended to help with the demographic goal by removing barriers to admission for applicants with low Graduate Record Exam (GRE) scores or with low undergraduate grade point averages (UGPA).(*1*) Although effects of the rubric approach on demographics were not directly evaluated due to lack of demographic data, more abstract statistical methods were used to evaluate whether there had been noticeable changes to admissions procedures.(*1*)

In this Comment I make three main points, some of which have already been discussed by prominent statisticians.(*6*)

1. Although the paper only tentatively concludes: "Overall, the results of this initial investigation are suggestive that our admission process did change…"(*1*), using prior knowledge of the intended results allows use of simple statistical tests that show conclusively that the rubric system specifically succeeded in reducing the UGPA/GRE barriers, increasing the fraction of applicants with low grades or test scores who were admitted.
2. Although the paper argues, based on prior literature, that this change will not have an impact on graduate outcomes, a more careful reading of that literature, including papers they cite, indicates that the admissions changes may raise further challenges for programs attempting to raise graduation rates.
3. A data-editing technique that is advocated and used in some of the analyses is likely to give biased results and thus should generally be avoided.

Evaluating effects on admissions of the procedure change

Young et al. (*1*) primarily use two methods to check if the new procedure is admitting different students than the old procedure. These methods are both generic in that neither uses prior knowledge of what the new methods are intended to change. Generic methods, while useful for many purposes, need low sensitivity to avoid getting false positive results from the many possible changes implicitly tested over the whole range of the distribution functions examined. These methods are not especially sensitive to the feature of prior interest- whether admissions have been opened up for students who would previously have been rejected due to low scores.

The simplest method used is a Kolgomorov-Smirnov test(*7*), which compares two cumulative distribution functions (CDFs) to test the null hypothesis that they represent random samples from the same population. This test was unable to reject the null at alpha=0.05 for the CDFs of the UGPA or either of the Physics and Quantitative GREs. (*1*) Young et al. are commendably scrupulous in avoiding making any claims based on this non-significant effect. (*1*)

Although the CDFs are not explicitly given, visual inspection of the smoothed probability density functions (PDFs) of the distributions of accepted applicants shown in Figures 5-7 indicates a much larger left tail (low scores) for each metric after the procedural change, precisely the changes that were intended *a priori* to reduce admissions barriers.(*1*) The data points are fairly visible, allowing quantitative reconstruction of the left tails. To systematically compare low-scorer admissions in the two groups, we may check what fraction of the old-system accepted students scored below the tenth percentile (17th lowest) of the rubric-admitted students. This cutoff includes enough students to matter but does not get into the meat of the distribution where individual points become too hard to discern in the figures. As nearly as I can judge from the figures, the total number of these low old-system points for the three metrics shown was 29, i.e. 3.9% of the total rather than 10%. The one-sided Fisher exact p-value for the increase (i.e. the probability that an increase at least that big would happen by pure chance) is less than 0.001. For the Quantitative GRE, Physics GRE, and UGPA taken separately the one-sided Fisher exact p-values were less than 0.001, 0.10, and 0.04, respectively.

These counting results confirm the visual impression from the smoothed PDFs that more low-scorers were admitted in the new system, as hoped, although for the Physics GREs taken separately the change would not be large enough to meet the conventional alpha=0.05 significance standard. It seems that most of the low-scorers admitted via the new system would not have been admitted in the old system, so that the new system almost certainly succeeded at opening up the admissions process substantially in the intended direction.

Modeling results and methods

The bulk of Young et al.'s (*1*) technical discussion looks at differences between random forest models of the different admissions procedures to see if changes show up in the models. Some model differences were observed. A model of the old system produced an area-under-curve (AUC) of 0.76 on test samples after training on the other data, about halfway between entirely unpredictive (AUC=0.5) and completely predictive (AUC=1.0). The model of the new system gave a testing AUC=0.63 on the full set of applicants, only about a quarter of the way toward full prediction. Young et al. (*1*) correctly emphasize that these models do not do very well at predicting admissions, especially after the change to a rubric-based system.

For a large subset of the newer applicants the rubrics actually used in the admissions decisions were available. Each of the rubrics is treated as a categorical rating. These variables were:

| Physics coursework | Initiative | Research: Dispositions |
| Math coursework | Perseverance | Achievement orientation |
| Other coursework | Fit: Research | Contributions to community |
| Academic honors | Fit: Faculty | Contributions to diversity |
| Research: Variety or duration | Research: Quality of work | General GRE |
| Conscientiousness | Research: Technical skills | Physics GRE |

Most of these rubrics, e.g. "achievement orientation", capture qualities that admissions committees try to take into account but that would not be available to standard quantitative models. Thus one would expect that a model with access to a complete set of rubric ratings would do dramatically better at predicting admissions than would models that have no access to most of the variables and that only see versions of others (e.g. grades) that have not yet gone through somewhat noisy and subjective evaluations. The rubrics are not only more complete but also are *downstream* of that initial evaluation and thus should be more closely tied to decisions. Nevertheless, the random-forest model based on these detailed metrics as evaluated and used by the admissions committee only increased the testing AUC of the rubric model to 0.67, even lower than the 0.76 for a model that used only limited unprocessed input data for the old system.

The surprising inability to model admissions decisions accurately even using all the rubrics available to the committee strongly indicates that other factors beyond the extensive list of rubrics were considered. Prior work from the same MSU group gives indications of some important factors missing from the rubric models.(*5*) Logistic regression models of admission controlling for UGPA and Physics GRE showed strong positive direct predictive coefficients for females and under-represented minorities. (*5*) These demographic factors were not included in the current models, including the rubric-based one. Young et al. (*1*) explain that since Michigan law forbids the direct use of such factors "To comply with this law, our university's admissions system collects limited demographic data and our department chose not to record the information that was available when evaluating applicants." (*1*)

In order to deal with all the models' low AUCs, the paper explores a modified analysis in which some points are removed before the random forest model is used. The paper states "…there are cases where applicants have similar physics GRE scores and GPA, yet one applicant is accepted while the other is not. Given that cases such as these might add challenges to modeling the data, removing such applicants might allow us to better characterize the general trends in the data." (*1*) This data-editing is problematic because removing points where a UGPA-GRE model is not predictive makes models that include those factors look better at predicting than they actually are. Such editing can obscure the need for other modeling factors that account for outcome differences that are not explained by UGPA or GREs. It undermines the key reason given for using the random forest method: "We choose to apply a classification machine learning model under this computational framing, specifically random …due to the lack of assumptions on the data and random forest's feature importances." (*1*) (Gelman(*6*) briefly discusses the problems with this procedure.) Similar problems could arise whenever such a data-dropping method is used before modeling a complicated real-world process in which the important predictors are not known *a priori*. The editing pre-supposes what factors are important and biases the model to focus on those factors.

Young et al.'s (*1*) inclusion of models of the edited data does, however, help reinforce the conclusion that factors outside the official rubrics must have played an important role in the admissions decisions. Relatively minor considerations outside the rubrics could degrade the AUC of any rubric-based model if many applicants are near the admit/reject boundary on important criteria included in the rubrics. Since even after removal of borderline cases the testing AUC only rose to 0.70 (*1*), the outside-the-rubric factors must have been quite important, changing decisions from those predicted by rubric-informed models even in non-borderline cases.

Prior literature on graduate performance predictors

The question of which variables help predict graduation is central to the clearly stated theme of the paper— the importance of increasing graduation rates.(*1*) Programs that adopt new admissions criteria will need to be mindful of whether the changes tend to make it easier or more challenging to maintain graduation rates. Young et al. (*1*) cite three publications to support their claim that traditional predictors have little value(*3, 4, 8*), so that de-emphasizing them will not create new challenges. A careful look at the papers cited and other relevant ones does not support that conclusion.

One Young et al. citation is to a thesis(*8*) that found no significant predictors of outcomes. Since that study was based on a single program it was subject to range restriction and collider stratification bias. With N=54, it had little chance of finding any effects. One other study(*9*), not cited, also found no conventionally statistically significant evidence of GRE predictive power in a design that should minimize collider stratification bias, but it concerned less quantitative fields and had N=32. Those results contrast with an uncited larger study (N=138) lacking collider stratification bias that showed correlation coefficients between 0.55 and 0.70 between GRE scores and measures of psychology graduate students' performance in learning and using quantitative social science methods.(*10*)

Two papers(*3, 4*) that Young et al. (*1*) cite are specifically on whether GREs help predict completion of physics PhD programs. Although those papers did claim at most points (the Supplement to (*4*) is an exception) that the GREs were not significantly predictive, subsequent papers have shown that their analyses relied on myriad statistical errors, including a biased imputation method for missing data, collider stratification bias, improper treatment of range restriction, improper treatment of collinearity, improper use of null hypothesis significance testing on small subsets, and other errors. (*11, 12*) Each error tended to obscure the predictive power of GREs. (*11, 12*)  Reanalysis using conventional methods showed that, holding UGPA constant, the Quantitative and Physics GREs together provided a factor of 3 or more in predicting the odds of graduation between the 90$^{th}$ percentile and the 10$^{th}$ percentile in the large domestic cohort studied.(*12*)

Young et al. (*1*) justify the use of the complicated random forest technique to look for changes in the admissions criteria by citing a paper(*13*) comparing various techniques for predicting which students did well or poorly in a graduate data science program. Young et al. (*1*) do not, however, mention the substantive results of that paper. According to the random forest results shown in its Table 4, three of the top four predictors of who would do poorly (bottom 20$^{th}$ percentile) were GREs. (*13*) Three of the top five predictors of who would do well (top 20$^{th}$ percentile) were GREs. (*13*) Thus while this reference does offer support for the utility of random forest predictive models in some circumstances, its substantive results are evidence tending against the underlying premise that dropping or de-emphasizing GREs will create no difficulties for improving graduate performance in fields involving quantitative analysis of data. (*13*)

The implication of the prior relevant work, especially that cited by Young et al. (*1*), is then that admitting more students with low scores will require compensating changes either to other admissions criteria or to program structure to maintain or improve graduation rates. Perhaps some such changes are already integrated into the new rubrics. Follow-up studies of outcomes could be quite useful.

Acknowledgments

I thank Jamie Robins, Miguel Hernán, and Andrew Gelman for their patience in discussing the statistical issues raised and Ellen Fireman for her editorial advice.

References

1. N. T. Young, N. Verboncoeur, D. C. Lam, and M. D. Caballero, Rubric-based holistic review represents a change from traditional graduate admissions approaches in physics. *Physical Review Physics Education Research* **19**, 010134 (2023).

2. C. Miller, and K. Stassun, A test that fails. *Nature* **510**, 303 (2014).

3. C. W. Miller, B. M. Zwickl, J. R. Posselt, R. T. Silvestrini, and T. Hodapp, Typical physics Ph.D. admissions criteria limit access to underrepresented groups but fail to predict doctoral completion. *Sci. Adv.* **5**, eaat7550 (2019).

4. M. Verostek, C. W. Miller, and B. Zwickl, Analyzing admissions metrics as predictors of graduate GPA and whether graduate GPA mediates Ph.D. completion. *Physical Review Physics Education Research* **17**, 020115 (2021).

5. N. T. Young, and M. D. Caballero, Physics Graduate Record Exam does not help applicants ``stand out''. *Physical Review Physics Education Research* **17**, 010144 (2021).

6. A. Gelman, Rubric-based holistic word salad represents a change from traditional approaches to publishing papers in physics. *Statistical Modeling, Causal Inference, and Social Science*. 2023.

7. F. J. Massey, The Kolmogorov-Smirnov Test for Goodness of Fit. *Journal of the American Statistical Association* **46**, 68 (1951).

8. T. Wilkerson, The relationship between admission credentials and the success of students admitted to a physics doctoral program, PhD thesis, University of Central Florida,  (2007).

9. Sealy L, Saunders C, Blume J, and C. R., The GRE over the entire range of scores lacks predictive ability for PhD outcomes in the biomedical sciences. *PLoS ONE* **14**, e0201634 (2019).

10. B. E. Huitema, and C. R. Stein, Validity of the GRE Without Restriction of Range. *Psychological Reports* **72**, 123 (1993).

11. M. B. Weissman, Do GRE scores help predict getting a physics Ph.D.? A comment on a paper by Miller et al. *Science Advances* **6**, eaax3787 (2020).

12. M. B. Weissman, Invalid Methods and False Answers: Physics Education Research and the Use of GREs. *Econ Journal Watch* **19**, 4 (2022).

13. Yijun Zhao, Q. Xu, M. Chen, and G. Weiss, Predicting Student Performance in a Master's Program in Data Science using Admissions Data. *Proceedings of The 13th International Conference on Educational Data Mining (EDM 2020)*, 325 (2020).